# Traceability and Provenance in Big Data Medical Systems

Richard McClatchey, Jetendr Shamdasani, Andrew Branson, Kamran Munir & Zsolt Kovacs
CCS Research Cents, FET Faculty, UWE Bristol, UK
Email: richard.mcclatchey@uwe.ac.uk

Giovanni Frisoni & members of the neuGRID and neuGRID For Users (N4U) Consortia
IRCCS Centro San Giovanni di Dio FBF, Brescia, Italy
Email: gfrisoni@fatebenefratelli.it

*Abstract*— **Providing an appropriate level of accessibility to and tracking of data or process elements in large volumes of medical data, is an essential requirement in the Big Data era. Researchers require systems that provide traceability of information through provenance data capture and management to support their clinical analyses. We present an approach that has been adopted in the neuGRID and N4U projects, which aimed to provide detailed traceability to support research analysis processes in the study of biomarkers for Alzheimer's disease, but is generically applicable across medical systems. To facilitate the orchestration of complex, large-scale analyses in these projects we have adapted CRISTAL, a workflow and provenance tracking solution. The use of CRISTAL has provided a rich environment for neuroscientists to track and manage the evolution of data and workflow usage over time in neuGRID and N4U.**

*Keywords; Provenance, Workflows, Biomedical Analysis, Grid Computing, Neuroimaging*

## I. Introduction

With the increasing volumes of digitized medical data and the consequent complexity of clinical research algorithms the need to provide validation and reproducibility of analysis results from Big Data sets demands the capture and collation of rising amounts of meta-data [1]. Scientific workflows are increasingly required to orchestrate research processes in medical analyses, to ensure the reproducibility of analyses and to confirm the correctness of outcomes [2]. In a collaborative research environment, where researchers use each others' results and methods, traceability of the data generated, stored and used must also be maintained. All these forms of knowledge are collectively referred to as 'provenance' information.

The availability of provenance information (history, evolution and usage for example) about a medical analysis is often as valuable for traceability as the results of the medical analysis itself [3]. In any system where there are many data-sets, and versions of algorithms operating upon those data-sets, particularly when the analysis is repetitively conducted potentially by collaborating teams of researchers, it is imperative to retain a record of who did what, to which sets of data, on which dates, and for what purpose as well as recording the results of the analysis process itself [4]. This information needs to be logged so that analyses can be reproduced or amended and repeated as part of rigorous research processes. All of this information, normally generated through the execution of workflows enables the traceability of the origins of data (and processes) and, perhaps more importantly, their evolution between different stages of their usage. Capturing and managing this provenance data enables users to query analysis information, automatically generate workflows and to detect errors and exceptional behaviour in previous analyses.

Provenance essentially means the history, ownership and usage of data and its processing in some domain of interest. For example, logging the processing of datasets in the study of MRI scans to determine biomarkers of the onset of Alzheimer's disease. The knowledge acquired from executing neuroimaging workflows must be validated using such stored provenance data. In health informatics emphasis has been placed upon the provision of infrastructures to support researchers for the purpose of data capture, image analysis, and the processing of scientific workflows and the sharing of results. This may include browsing data samples and specifying and executing workflows (or pipelines) of algorithms required for neurological analysis. To date none have considered how such analysis can be tracked over time, between researchers and over varying data samples and analysis workflows. This paper addresses that deficiency.

The dynamic and geographically distributed nature of Cloud computing makes the capturing and processing of provenance information a major research challenge. To date provenance gathering systems and techniques have mostly been based solely on workflows within scientific research domains such as bioinformatics (e.g. [5], [6]); for further discussion on provenance the interested reader is directed to [7]. However existing state-of-the-art provenance management systems are not generic and reconfigurable "on-the-fly". Most workflow provenance management services such as LONI [8] are designed only for data-flow oriented workflows and researchers are now realising that tracking data alone is insufficient to support the scientific process (see [5]). In this article we outline the provenance management approach developed in the neuGRID [9] and neuGRIDforUsers (N4U) projects to preserve the data collected in the execution of neuroimaging analysis workflows.

These projects have adopted an 'on-the-fly' reconfigurable tracking system, called CRISTAL [10] that was produced at CERN CMS [11], to provide provenance management in tracking neurological analyses. Due to its reconfigurable nature CRISTAL was ideally adapted to managing data-flow oriented workflows and control flows in the neuGRID and N4U projects.

This article proceeds as follows: Section II introduces the N4U Virtual Laboratory and its Analysis Base. Section III describes the use of CRISTAL in the neuGRID and N4U projects. The Analysis Service developed in neuGRID/N4U is expanded in Section IV and section V discusses future research directions.

## II. THE N4U VIRTUAL LABORATORY

Research in computational infrastructures for Alzheimer's imaging analysis includes neuGRID [9], NeuroLog [12], CBRAIN [13], BIRN [14] and LONI [8]. In these efforts data gathering, management and visualization has been facilitated, and the constituent data is captured and stored in large distributed databases. Such data management makes it unrealistic for clinical researchers to constantly analyze dynamic and huge data repositories for research. In future, it is highly likely that data volumes and their associated complexities will continue to grow, especially due to the increasing digitization of (bio-) medical data. Therefore, users need their data to be more accessible, understandable, usable and shareable, as in N4U.

In N4U project we have provided a Virtual Laboratory (VL, see https://neugrid4you.eu) which offers neuroscientists tracked access to a wide range of datasets, algorithm applications and to computational resources and services in their studies of biomarkers for identifying the onset of Alzheimer's disease. The N4U VL, whose architecture is illustrated in Figure 1, is based on services layered on top of the neuGRID infrastructure. It was developed for imaging neuroscientists involved in Alzheimer's studies but has been designed to be generically reusable across other medical research communities.

The N4U VL has been designed to provide access to infrastructure-resident data and to enable the analyses demanded by the biomedical research community. This VL enables clinical researchers to find clinical data, pipelines, algorithm applications, analysis definitions and detailed interlinked provenance in an intuitive environment. This has been achieved by basing the N4U VL on an integrated Analysis Base [15] (as depicted in figure 1), which has been developed following the detailed user requirements in N4U. The high-level flow of data and analysis operations between various components of the virtual laboratory and the analysis base are highlighted in Figure 1. The analysis base supports processing by indexing and interlinking the neuroimaging and clinical study datasets stored on the N4U Grid infrastructure, algorithms and scientific workflow definitions along with their associated provenance information (so-called 'provenance enabled objects). Once researchers conduct their analysis using this interlinked information, the analysis definitions and resulting data along with the user profiles are made available in the analysis base to researchers for tracking and reusability purposes via an Analysis Service. In overview the N4U virtual laboratory comprises the following components (1) Information Services; (2) Analysis Services; (3) The Analysis Service Work Area; and (4) Science Gateway and Specific Support Centre. The Information Services comprise the Persistency and Querying services within the N4U virtual laboratory. The Persistency Service provides appropriate interfaces for storing the meta-data of datasets, such as ADNI [16], in the analysis base such that these datasets become indexed in the base and available for use.

The N4U Analysis Service provides access to tracked information (images, pipelines and analysis outcomes) for querying/browsing, visualization, pipeline authoring and execution. Its Workarea is a facility for users to define new pipelines or configure existing pipelines to be run against selected datasets and to be dispatched to conduct analyses. The N4U Science Gateway in the VL provides facilities that include a Dashboard (user interface), Online Help and service interfaces for users to interact with the underlying set of N4U services. It is currently in use and under evaluation by research clinicians in N4U sites at Fatebenefratelli Brescia, Univ. Hospital Geneva, VUmc Amsterdam and Karolinska Institute, Stockholm.

## III. PROVENANCE TRACKING : THE CRISTAL SYSTEM

CRISTAL is a data and process tracking system that was developed to support the construction of the CMS detector at the CERN LHC [11]. The software is mature having been used to provide 24/7 operation over a decade with minimal maintenance is scaleable over 100,000s of elements, curating over a Terabyte of data. Using the facilities for description and dynamic modification inherent in CRISTAL in a generic and reusable manner, we have provided dynamically modifiable and reconfigurable workflows and data set definitions for N4U.

The so-called "description-driven" nature of CRISTAL models allows dynamic actions on process instances already running, and users can even intervene in the actual process instances during execution (for further detail refer to [10]). Thus the clinical workflows ('pipelines') can be dynamically (re)-configured based on the context of execution without compiling or stopping overall processing and the user can make modifications directly upon any process parameter, whilst preserving all historical versions so they can run alongside the new version. This is a genuinely novel feature for clinical research systems. Consequently in neuGRID/N4U, we have used CRISTAL to provide the provenance needed to support neuroscience analysis and to track individualized analysis definitions and usage patterns, thereby creating a practical and queryable knowledge base for neuroscience researchers. The N4U Analysis Base thus manages the accumulated project analysis records and supports the collaborative, verifiable and reproducible research needed to support N4U's studies of Alzheimer's biomarkers.

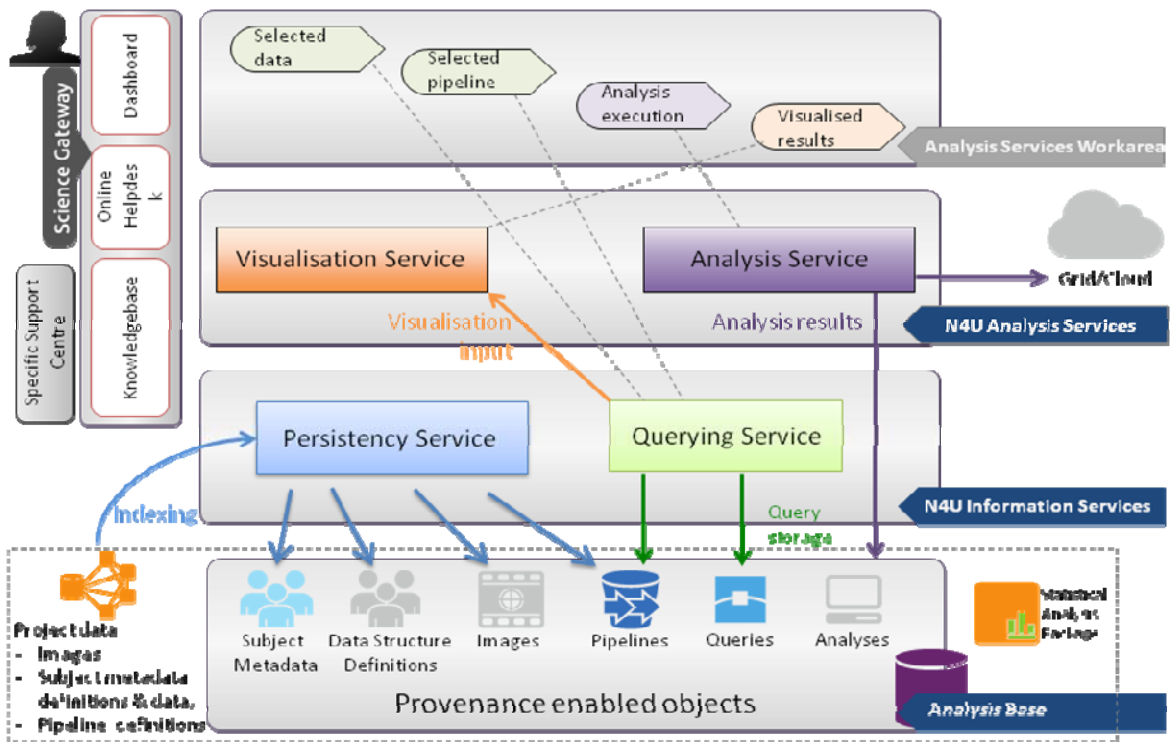

**Figure 1:** The N4U Virtual Laboratory

CRISTAL captures the provenance data resulting from the specification and execution of the stages in analysis workflows. Its provenance management service also keeps track of the origins of the data products generated in an analysis and their evolution between different stages of research analysis. CRISTAL records every change made to its objects, which are referred to as *Items* and can be data, process or agent elements (atomic or composite) each with a lifecycle. Whenever a modification is made to any Item, the definition of that Item or its application logic, the modification and the metadata associated with that change (e.g. who made the change, when and for what purpose) are stored in CRISTAL alongside that data as Items themselves. Importantly this makes CRISTAL applications fully traceable and easy to manage, and this data may be used as provenance information. In N4U, CRISTAL manages data from the Analysis Service, containing the full history of computing task execution; it can also provide this level of traceability for any piece of data in the system, such as the datasets, pipeline definitions and queries.

Provenance querying facilities are provided by the Querying Service in neuGRID/N4U (see figure 1). Users can retrieve past analyses, retrieve specific versions of a workflow and examine the results of each individual computation and thereby track usage and ownership of workflows. A key ability of the CRISTAL system is its ability to adapt to changing requirements in terms of provenance storage. The domain of neuroscience is constantly changing as new workflows, algorithms and research studies are developed. The underlying CRISTAL model allows the system to evolve to handle such changes whilst retaining provenance information in a consistent and traceable manner.

In execution CRISTAL tracks the following information:
- Workflow specifications;
- Data or inputs supplied to each workflow component;
- Annotations added to the workflow and its components;
- Links and dependencies between workflow components;
- Execution errors generated during analysis and
- Outputs produced by each workflow component.

At the heart of the neuGRID/N4U infrastructure is a distributed computation environment designed to efficiently handle the running of image processing workflows such as the cortical thickness-measuring algorithm, CIVET [17]. This is not enough on its own, however, as users require more than simply raw processing power. They need to be able to access a large distributed library of data and to search for a group of images with which they want to work. A set of common image processing workflows is also necessary within the infrastructure. A significant proportion of clinical research involves the development of customized workflows and image analysis techniques. The ability to edit existing scientific workflows on-the-fly and to construct new workflows using established tools is essential and novel in N4U. Researchers need to be able to examine each stage in the processing of an analysis workflow in order to confirm that it is verifiably accurate and reproducible. Part of the CRISTAL data model is shown in figure 2.

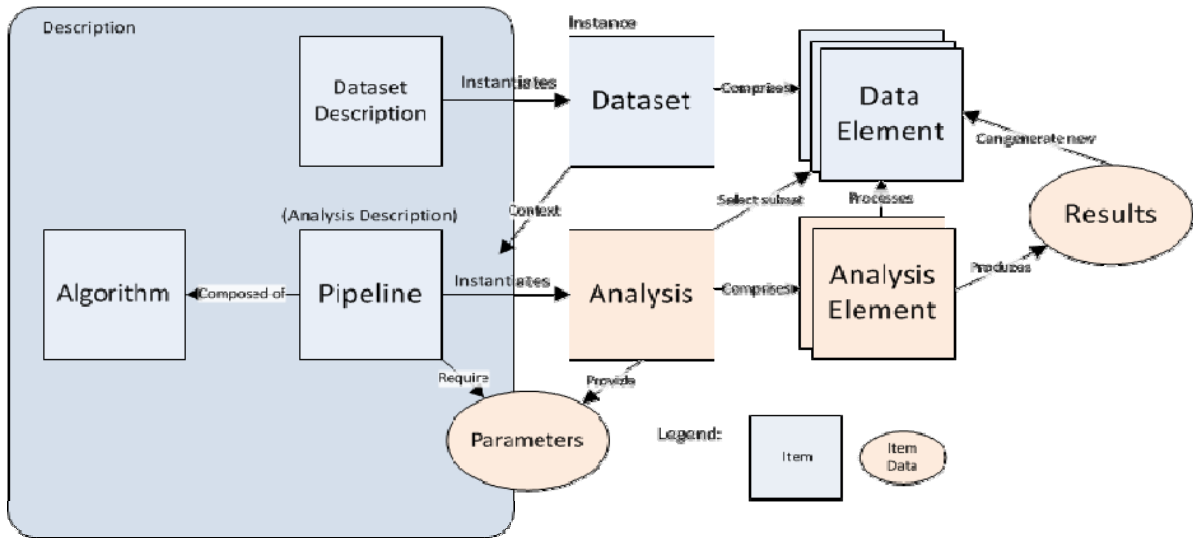

**Figure 2:** Detail of the CRISTAL Model including Items such as Data and Analysis Elements.

The figure shows CRISTAL *Items* (i.e. description-driven objects either descriptions or instances of descriptions) as squares and external project data as ovals. Objects associated with user analyses are shown in orange. The model shows how pipelines and datasets are used to create parametrized analyses that produce results as outcomes. The figure also shows objects called Data Elements and Analysis Elements which provide the substance of all the Provenance Enabled Objects stored in the N4U Analysis Base (as shown in Figure 1). The model contains *Items* holding metadata on all the pipelines and datasets registered in N4U.

Pipeline *Items* give the location of the analysis scripts that need be run, along with default execution environment settings, and any common directory locations that should be passed with the job. Dataset *Items* contain Data Elements, which are sets of files that should be processed together in one job, along with specific metadata about that set that can vary in composition between different datasets [18].

The central *Item* type of the Analysis Service itself is the 'Analysis' object (figure 2), which is a user-initiated process comprising the execution of a pipeline on one or more elements of a single dataset. Analyses are instantiated from Pipeline *Items*, with a Dataset and some Data Element IDs identified as parameters. Pipeline *Items* can require additional parameters that may be given when the user creates the Analysis to override or append the defaults included in the Pipeline specification. Each Analysis belongs to the user that created it, and can only be viewed by that user, this information also being logged by CRISTAL. In execution the Analysis suite instantiates an Analysis Element for each given Data Element creating an instance of the Pipeline workflow that can be dispatched to the Grid as executable jobs and whose provenance is gathered for each step of the workflow execution for tracking by CRISTAL.

IV. THE ANALYSIS SERVICE IN OPERATION

The N4U Persistency Service stores clinical datasets in the Grid and indexes their metadata in the Analysis Base (see figure 1). This is the starting point for a user's interaction with the Analysis Service. To create a new analysis, the user must browse the Analysis Base to decide which data she wants to analyze, and which analysis algorithm(s) she wants to be run on it. These choices essentially define the context for the user's analysis and need to be stored as a defining record of that analysis, potentially along with user annotation. Optionally, the user may add post-processing work to the analysis to generate visualizations or summary reports after the algorithm has run.

The Analysis Service provides *workflow orchestration* for scientists and a platform for them to execute their experiments on the Grid. It allows users to recreate their experiments on the neuGRID/N4U Infrastructure using previously recorded provenance information and to view their results via visualization tools and to perform statistical analyses. It enables:

- The browsing of past analyses and their results;
- The creation of new analyses by pairing datasets with algorithms and pipelines found in the Analysis Base;
- The execution of analyses by creating jobs to be passed to the Pipeline Service, then logging the returned results in the analyses objects;
- Re-running of past analyses with different parameters or altered datasets and
- The sharing of analyses between researchers.

Once the specification of an analysis is complete, it can be run. This involves sending jobs out to the Grid for every element of the analysis algorithm selected, for every element of the selected dataset. This is done by creating a child analysis for each dataset element, with its own instance of the algorithm to be run, from which the Grid jobs are derived. This could potentially generate a large amount of parallel work, which the Grid will distribute to many computing elements. As the elements of the child analysis complete, the Analysis Service keeps records of the state of the result set, which Grid resources were used for the computation, and the exact times the operation was started and completed as part of the provenance data for that analysis. If a pipeline was selected that requires more than one computational step to complete, or allows for a particular analysis to fork its work into more than one thread, then more jobs will be sent out until the pipeline is complete. Each of these is recorded, again as the provenance data of the analysis for subsequent usage.

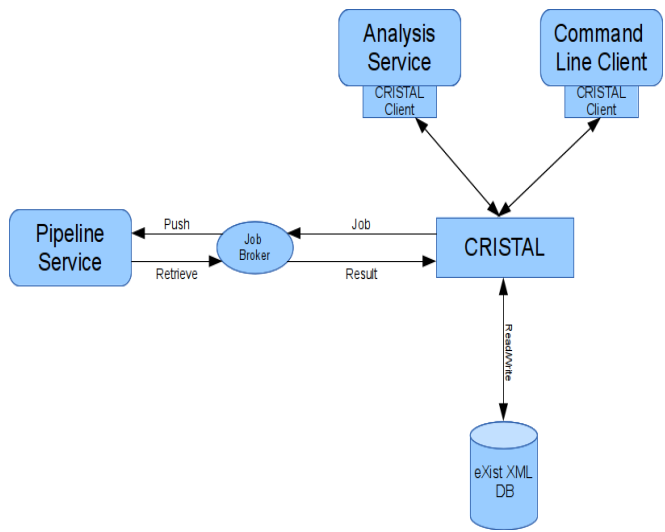

**Figure 3:** The N4U Analysis Service and CRISTAL

Once the child analyses of each dataset element are complete, any specified post-processing steps of the whole analysis can be performed in order to aggregate the results, generating tables and figures with external software packages from the data produced by the pipelines and the provenance of that result data. All of the data resulting from the analysis and its provenance is stored in the Analysis Base, where it enriches the dataset and algorithm profiles with real-world usage data essentially creating a knowledge base for users to query. An existing, complete analysis may be cloned, its parameters and subjects altered if desired, and run again. All analysis objects belong to, and are by default only visible to, the user who created them. It is possible to share analyses between users in the project; this functionality will be enhanced with more social-network style functionality in the final stages of the project.

The detailed operation of the Analysis Service is best understood with a practical example. Consider the case where a clinician wishes to conduct a new analysis. Her first step would be to compile a selection of data from the datasets which are available to her. To do this she would log into the Analysis Service Area and interact with the Querying Service through its user interface to find data that possesses the particular properties she is looking for ((see figure 1). She submits her constraints, which are passed as a query to the Querying Service. The Querying Service then queries the Analysis Base which would return a list of dataset properties and locations which meet her constraints. The Querying Service interface would then display this list to the clinician to approve. Once the user is satisfied with her dataset selection she combines it with a pipeline specification to create her Analysis. To do this she would need to use the Analysis Service Interface to search CRISTAL for existing algorithms that she can use to create a new pipeline or select a pre-defined pipeline. Command line utilities will be provided to aid in the creation of a pipeline by connecting different algorithms together as steps. The completed pipeline will have a dataset associated with it and once this pipeline is ready it will be run on each element of the dataset by CRISTAL.

The pipeline will be sent to CRISTAL which will orchestrate the input pipeline (see figure 3) using a Job Broker and the N4U Pipeline Service. Single activities from the input workflow will be sent to the Pipeline Service as a single job using the pipeline API. Once the job has completed, the result will be returned to CRISTAL. Here CRISTAL will extract and store provenance information for this job. This information will contain factors such as the time taken for execution, and whether the job completed successfully. It will store this information internally in its own data model. It will also post this information to the Analysis Base so that this crucial provenance information is accessible by the Querying Service. This loop of sending jobs and receiving the result will continue until the workflow is complete. Once this workflow has completed CRISTAL will once more generate provenance information and store this provenance for the entire workflow in its own internal data store and the Analysis Base. The final result of the completed workflow/pipeline will be presented to the user for evaluation. A link to the completed result in the form of a LFN (a GRID location) will be stored in the Analysis Base.

The clinician now has a permanently logged record (provenance data) of her analysis including the datasets and (versions of) algorithms she has invoked, the data captured during the execution of her analysis and the final outcome and data returned by her analyses. These provenance elements may also have associated annotation that she has added to provide further knowledge of her analysis that she or others could consult at a later time to re-run, refine or verify her analysis.

## V. CONCLUSIONS AND FUTURE DIRECTIONS

We have outlined the neuGRID/N4U approach that has been developed to capture and preserve the provenance data that emerges in the specification and execution of (stages in) analysis workflows, and in the definition and refinement of data samples used in studies of Alzheimer's disease (AD). In the neuGRID/N4U projects a service has been implemented that captures workflow information in a scaleable, project-wide provenance database from data gathered in the execution of scientific workflows. This database keeps track of the origins of the data and its evolution between different stages of research analysis and allows users to query analysis information, to regenerate analysis workflows, to detect errors and to reproduce and validate analyses. The management of provenance data has been based on the CRISTAL software (see [18]), which is a data and workflow tracking system. It addresses the harmonization of processes by the use of a kernel, so that potentially multiple heterogeneous processes can be integrated with each other, and have their workflows tracked. Using the facilities for description and dynamic modification in CRISTAL in a reusable manner, neuGRID/N4U is able to provide modifiable and reconfigurable workflows for a wide variety of healthcare applications.

In the future we intend to research and develop a so-called User Analysis module that will enable applications to learn from their past executions and improve and optimize new studies and processes based on the previous experiences. Models will be formulated that can derive the best possible optimisation strategies by learning from the past execution of experiments and processes. These models will evolve over time and will facilitate decision support in designing, building and running the future processes and workflows in a domain. A provenance analysis mechanism will be built on top of the data that has been captured in CRISTAL. It will employ approaches to learn from the data that has been produced, classify and reason from the information accumulated and present it to the system in an intuitive way. This information will be delivered to users while they work on new workflows and will be an important source for future decision-making. One essential future element is the provenance interoperability aspect present within the neuGRID/N4U projects. Currently, we are working on exporting the provenance enabled objects to the emerging PROV [19] interoperability standard. This will allow N4U users to use their provenance data in other PROV-compliant systems.

There are further plans to enrich the CRISTAL kernel (the data model) to model not only data and processes (products and activities as Items) but also to model agents and users of the system (whether human or computational). We will investigate how the semantics of CRISTAL items and agents could be captured in terms of ontologies and thus mapped onto or merged with existing ontologies for the benefit of new domain models. The emerging technology of cloud computing and its application in complex domains, such as medicine and healthcare, provide further interesting challenges, particularly for healthcare.

ACKNOWLEDGEMENTS

The authors acknowledge the financial support of the EC through the Grant Agreement numbers 211714 & 283562. They also thank their projects partners: FBF (Brescia, Italy), UWE (Bristol, UK), Gnubila (Annecy, France), VUmc (Amsterdam, NL), HealthGrid (Lyon, France), Prodema (Braunshofen, Switzerland), Univ. Hospital Geneva (Switzerland), CEA (Paris, France), CF Consulting (Milan,. Italy) and Karolinska Institute (Stockholm, Sweden).

VI. REFERENCES

[1] F. Estrella et al., "Experiences of Engineering Grid-Based Medical Software", International Journal of Medical Informatics, Vol. 76, No. 8 pp 621-632 Elsevier publishers. August 2007.

[2] Y. Gil et al., "Examining the challenges of scientific workflows," Computer, vol. 40, pp. 24-32, 2007.

[3] S. Miles et al., "Provenance: The bridge between experiments and data," Computing in Science & Engineering, vol. 10, pp. 38-46, 2008.

[4] A. Dolgert et al., "Provenance in high-energy physics workflows," Computing in Science & Engineering, vol. 10, pp. 22-29, 2008.

[5] S. Bechhofer et al., "Why Linked Data is not Enough for Scientists". Future Generation Computer Systems Vol 9 No. 2 pp 599-611, Elsevier Publishers, 2013.

[6] S. Davidson et al., "Provenance in Scientific Workflow Systems". IEEE Data Engineering Bulletin. 30(4): 44-50, 2007

[7] Y. Simmhan et al., "A Survey of Data Provenance in e-Science". In SIGMOD RECORD, Vol 34, P. 31-36. ACM, 2005.

[8] M. Pan et al., "The LONI Workflow Processing Environment: Improvements for Neuroimaging Analysis Research," 11th Annual Meeting of the Organization for Human Brain Mapping, 2005.

[9] A. Redolfi, et al., (2009). "Grid infrastructures for computational neuroscience: the neuGRID example". Future Neurology, Vol 4 No. 6, pp. 703-722 Future Medicine Ltd, London, 2009.

[10] A. Branson et al., "CRISTAL : A Practical Study in Designing Systems to Cope with Change". Information Systems 42, pp 139-152. Elsevier publishers

[11] S. Chatrchyan et al., "The CMS at the CERN LHC. The Compact Muon Solenoid Collaboration", The Journal Instrumentation Volume: 3 Article No: S08004, Institute of Physics Publishers 2008

[12] J. Montagnat et al., "NeuroLOG: a community-driven middleware design," HealthGrid 2008, Chicago, 2008.

[13] CBRAIN Project, http://cbrain.mcgill.ca/. Accessed January 2015.

[14] J. S. Grethe et al., 2005, "BIRN: Biomedical informatics research network". Stud Health Technol Inform. 2005;112:100-9.

[15] K. Munir et al., "An Integrated e-Science Analysis Base for Computational Neuroscience Experiments & Analysis". Procedia - Social & Behavioral Sciences. Vol 73 pp 85-92.

[16] S. G. Mueller, M. W. Weiner. L. J. Thal, et al., "The Alzheimer's disease neuroimaging initiative", Neuroimaging Clinics of North America, vol 15, no 4, pg. 869, 2005, NIH Public Access.

[17] A. P. Zijdenbos et al., "Automatic "pipeline" analysis of 3-D MRI data for clinical trials: application to multiple sclerosis," IEEE Transactions on Medical Imaging, vol. 21, pp. 1280-1291, 2002.

[18] R. McClatchey et al, "Providing Traceability for Neuroimaging Analyses". International Journal of Medical Informatics, 82 pp 882-894, Elsevier Publishers, 2013.

[19] P. Groth & L. Moreau "An Overview of the PROV Family of Documents". Details availablfr the W3 Consortium at: www.w3.org/TR/2013/NOTE-prov-overview-20130430/ Accessed January 2015.